\documentclass[conference]{IEEEtran}
\usepackage{fancyhdr}
\usepackage{cite}
\usepackage{amsmath,amssymb,amsfonts}
\usepackage{algorithmic}
\usepackage{graphicx}
\usepackage{textcomp}
\usepackage{booktabs}
\usepackage{makecell}
\usepackage[table]{xcolor}
\def\BibTeX{{\rm B\kern-.05em{\sc i\kern-.025em b}\kern-.08em
    T\kern-.1667em\lower.7ex\hbox{E}\kern-.125emX}}
\usepackage{url}
\usepackage{subcaption}

\usepackage{listings}

\lstdefinestyle{yaml}{
  basicstyle=\ttfamily\scriptsize,
  columns=fullflexible,
  keepspaces=true,
  showstringspaces=false,
  breaklines=true,
  breakatwhitespace=false,
  postbreak=\mbox{\textcolor{gray}{$\hookrightarrow$}\space},
  frame=single,
  framerule=0.3pt,
  framesep=3pt,
  xleftmargin=2pt,
  framexleftmargin=2pt,
  xrightmargin=2pt,
  aboveskip=2pt,
  belowskip=2pt,
  commentstyle=\color{green!40!black}\itshape,
  stringstyle=\color{red!70!black},
  keywordstyle=\color{blue!70!black}\bfseries,
  keywordstyle=[2]\color{blue!50!red!80!black}\bfseries,
  morecomment=[l]{\#},
  morestring=[b]",
  morestring=[b]',
  morekeywords={
    site_id,evidence_id,h100,
    slurm,partitions,gpu_count_per_node,
    storage,scratch,path_pattern,compute_visible,
    evidence,id,source_type,field,value,
    evidence_ref,conflicting_refs,source_ref,
    url,heading,chunk_id,quote,pilot_id
  },
  morekeywords=[2]{true,false,null}
}

\begin{document}

\title{The Other Half of Workflow Portability: Evidence-Backed HPC Site Profiles\\with Agentic Discovery}

\author{
\IEEEauthorblockN{Md Saiful Islam, Douglas Thain}
\IEEEauthorblockA{\textit{University of Notre Dame}, Notre Dame, IN, USA\\
\{mislam5, dthain\}@nd.edu}
}
\maketitle

\thispagestyle{fancy}
\lhead{}
\rhead{}
\chead{}
\lfoot{\parbox{\textwidth}{\scriptsize
\copyright~2026 IEEE. Personal use of this material is permitted. Permission from IEEE must be obtained for all other uses, in any current or future media, including reprinting/republishing this material for advertising or promotional purposes, creating new collective works, for resale or redistribution to servers or lists, or reuse of any copyrighted component of this work in other works.}}
\rfoot{}
\cfoot{}
\renewcommand{\headrulewidth}{0pt}
\renewcommand{\footrulewidth}{0pt}

\begin{abstract}
Moving a workflow developed and tested at one HPC site to another rarely succeeds without some amount of trial and error. Package managers rebuild software environments, containers ship whole filesystems, and workflow specifications such as backpacks package a workflow with its software, data, and resource requirements. These approaches address one half of workflow portability: what a workflow needs. But none describes how a given HPC site must be used, and that missing half is why even a portable workflow requires manual adjustment at each new site. That gap includes the site's resource shape, storage configuration, network permissions, and operating policies. This information may be explicit in the batch system, hidden in the prose of documentation, or buried deep within a router's configuration, making it difficult for an automated deployment tool to turn site knowledge into useful deployment decisions. We propose the HPC site profile, a structured, evidence-backed document that makes this knowledge actionable. We automatically construct it in three steps that mirror where the information lives: measuring the login node, extracting typed fields from documentation with a bounded language-model agent, and submitting pilot jobs for eligible unresolved fields. Every field is verified against its evidence or discarded, so a rule, not the model, decides what enters the profile. The profile then preflights a workflow into an execution plan or an early, explainable failure. We build profiles at Purdue Anvil, TACC Stampede3, and Notre Dame CRC and present a case study of preflighting a real workflow.
\end{abstract}

\begin{IEEEkeywords}
scientific workflows, workflow portability, high performance computing, information retrieval, RAG, agentic AI
\end{IEEEkeywords}

\section{Introduction}

Scientific workflows are becoming increasingly portable and are shared more
often across research groups and HPC sites. Package managers reconstruct complex 
software environments from a specification alone~\cite{gamblin2015spack,anaconda_conda_pack_2018}.
Containers carry an entire filesystem image, so that an application arrives at
a new machine with its dependencies already resolved~\cite{priedhorsky2017charliecloud, kurtzer2017singularity}. 
Backpacks extend the same idea beyond software, capturing a workflow together 
with its software, data, and compute specifications in a single portable
artifact~\cite{floability-backpack-escience25}. Each of these specifies what a 
workflow needs, yet none describes how a given HPC site must be used, 
so deploying at an unfamiliar site still takes some manual adjustment.

\begin{figure}[t]
  \centering
  \includegraphics[width=0.45\textwidth]{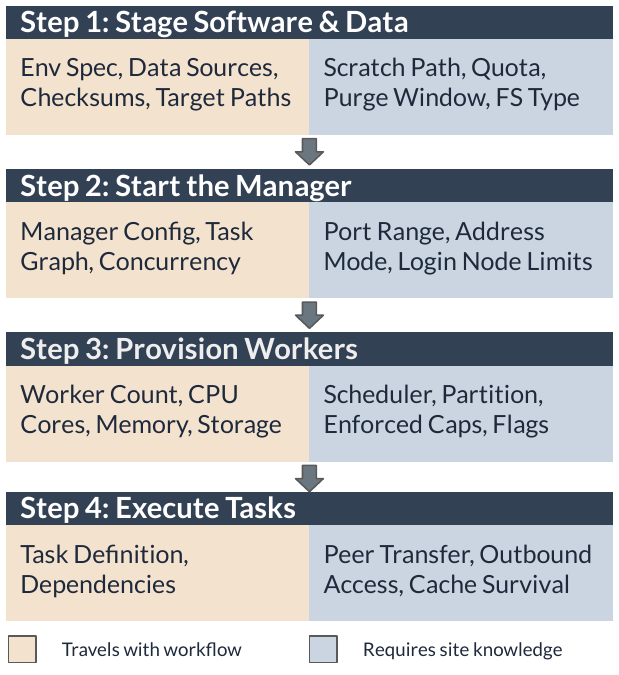}
  \caption{A typical workflow deployment proceeds in four steps. At each step,
some decisions travel with the workflow (left), while others require
site-specific knowledge (right). Packaging tools handle the former, but the
latter still forces manual tuning at every new site.}
  \label{fig:deployment}
  \vspace{-1.6em}
\end{figure}

A portable workflow deployment requires more than what the workflow carries.
Figure~\ref{fig:deployment} decomposes a typical deployment
into four steps and separates, at each one, the decisions that travel
with the workflow from those that can only be answered by the site.
Before a single task runs, one must know which batch scheduler the site uses,
which partition to select, which submission flags are required, where data may
be staged, and whether a worker process on a compute node may open a connection 
back to a manager or to another worker on a different node. These values and permissions differ
across sites, and portable workflows provide no source from which to obtain them. Users are therefore left to 
determine the values  themselves, and often rely on conventional defaults: across three
systems, the most common source of walltime overestimation was the selection of a default 
value such as an hour or a day~\cite{brown2022predicting}. Some incorrect values are 
rejected immediately, but others are discovered only after a job has waited in the 
queue and its software and data have been prepared, wasting both time and allocation.

The knowledge needed to close this gap is scattered among fragmented sources. 
Some is visible to a login-node command, though that command can mislead and report a limit the site does not enforce. Some is stated only in a user guide or FAQ. And some can be learned only by running a job and watching what happens. Assembling a complete picture means consulting all of these sources and reconciling them where they disagree. Beyond operational friction, these undocumented constraints feed the computational reproducibility crisis, impeding independent verification across facilities~\cite{antunes2024reproducibility,da2024workflows}. Our prior work is a case in point: the backpack made workflows substantially more portable~\cite{floability-backpack-escience25}, yet required manual adjustment at every new site.

To close this missing half, we propose an \emph{HPC site profile}, a
structured, machine-readable description of a site whose every field can be
traced to supporting evidence. A user could construct such a profile by reading
the site's documentation, running commands on the login node, and submitting
small test jobs. Creating a profile manually would be sufficient to bridge the
site-knowledge gap. However, the process must be repeated for every new site
and revisited as resources and policies change. Even a correct profile can become 
stale when documentation is updated incompletely or
falls out of date~\cite{tan2024detecting}.

We therefore automate profile construction in three steps. First, we probe the login node. Second, we extract typed fields from the site's authoritative documentation. Third, we submit short pilot jobs to observe compute-node properties and test network policy. We delegate the second step to a language model. A bounded, read-only agent gathers the documentation, and a separate extraction stage proposes fields with citations. Every field is then verified deterministically or discarded. Eligible unresolved fields become work orders for the pilot jobs, while the others remain unset. After the profile is built, the tool can preflight a workflow and return either an execution plan grounded in site evidence or an early, explainable failure before submission.

We build profiles for Purdue Anvil~\cite{song2022anvil}, TACC
Stampede3~\cite{stampede3}, and the Notre Dame CRC~\cite{ndcrc}, which span two
scheduler families, and preflight a real workflow packaged as a
backpack~\cite{floability-backpack-escience25}. We evaluate 108 profile builds
covering three sites, four models, three retrieval strategies, and three
repetitions. The resulting profiles are compared against hand-written ground
truth using field accuracy, precision, recall, abstention behavior, and citation
validity. We also compare model and retrieval choices by token cost, latency,
and run-to-run variance, and report where documented and measured values
disagree.

This paper makes the following contributions:
\begin{enumerate}
\item The \emph{HPC site profile}, a structured, evidence-backed description of how a site must be used and what behavior it enforces.
\item A three-step construction pipeline combining login-node measurement, documentation extraction, and pilot-job probing, with verifiable evidence for every field.
\item A preflight step that produces an execution plan or an early, explainable failure before submission.
\end{enumerate}

\section{The Missing Half}
\label{sec:missing}

\begin{figure}[t]
\centering
\begin{minipage}{0.98\columnwidth}
\begin{lstlisting}[style=yaml]
site_id: tacc-stampede3
evidence_id: 7f3a2c
slurm:
  partitions:
    h100:
      gpu_count_per_node:
        value: 4
        evidence_ref: doc-89
        conflicting_refs: measure-04
storage:
  scratch:
    path_pattern:
      value: "/scratch/10724/{username}"
      evidence_ref: measure-17
    compute_visible:
      value: true
      evidence_ref: pilot-03
# ... other site-profile fields omitted
\end{lstlisting}
\end{minipage}
\small{(a) Constructed site profile}

\vspace{0.7em}

\begin{minipage}{0.98\columnwidth}
\begin{lstlisting}[style=yaml]
site_id: tacc-stampede3
evidence_id: 7f3a2c
evidence:
- id: measure-04
  source_type: measurement
  field: "/slurm/partitions/h100/gpu_count_per_node"
  value: null
  source_ref: login-measurements.json
- id: doc-89
  source_type: documentation
  field: "/slurm/partitions/h100/gpu_count_per_node"
  value: 4
  url: "docs.tacc.utexas.edu/hpc/stampede3/"
  heading: "GPU Nodes > H100 > Table 5a"
  chunk_id: doc-hpc-stampede3:c25
  quote: "GPU: | 4x NVIDIA H100 SXM5"
- id: measure-17
  source_type: measurement
  field: "/storage/scratch/path_pattern"
  value: "/scratch/10724/{username}"
  source_ref: login-measurements.json
- id: pilot-03
  source_type: pilot
  field: "/storage/scratch/compute_visible"
  value: true
  pilot_id: 0d57a2458f4b450b
# ... other evidence entries and links omitted
\end{lstlisting}
\end{minipage}
\small{(b) Referenced evidence report}

\caption{Linked excerpts from a Stampede3 site profile build and its evidence report. Shared identifiers tie each profile value to measurement, documentation, or pilot evidence. The profile uses the documented GPU count because Slurm reports no value. It takes the scratch path from
measurement and records pilot-confirmed compute visibility.}
\label{fig:profile-excerpt}
\vspace{-1.2em}
\end{figure}

\subsection{What Deployment Needs from a Site}

As shown in Figure~\ref{fig:deployment}, deploying a manager-worker workflow
requires several site-specific decisions that do not travel with the workflow.
Staging software and data depends on available filesystems,
compute-node visibility, quotas, and retention policies. Starting the manager
depends on whether login-node processes are permitted, which address to
advertise, and which ports may be used. Provisioning workers requires the
scheduler, partition, account, walltime, and site-specific submission options.
Task execution may further depend on worker-to-worker communication and external
network access. Without this knowledge,
a workflow manager may choose inefficient defaults or fail altogether.

\begin{figure*}[t]
  \centering
  \includegraphics[width=0.95\textwidth]{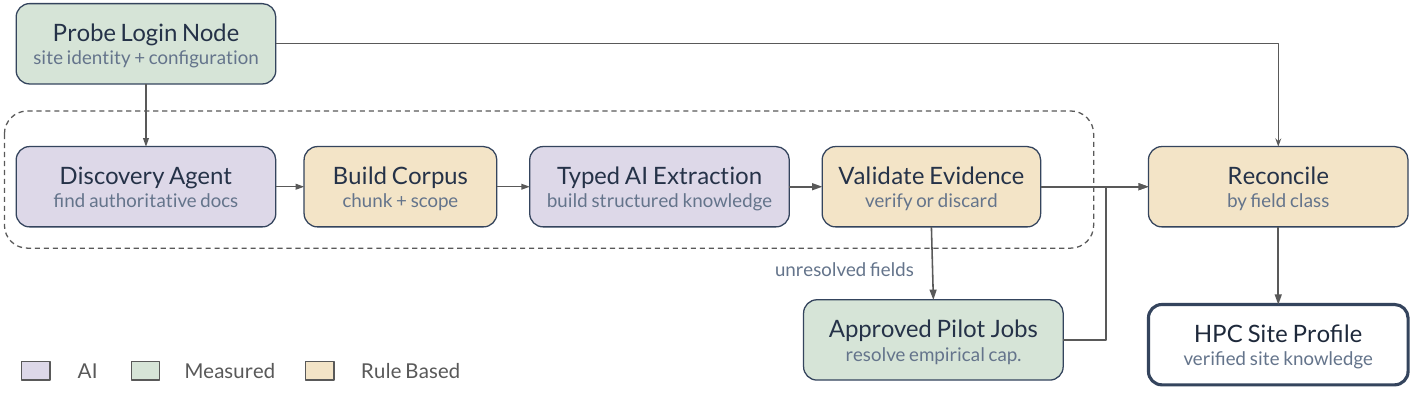}
\caption{Constructing an HPC site profile. Login-node measurements scope
documentation queries, but measured values are withheld from extraction so
documented values are proposed independently. Within the documentation path
(dashed), every proposed field is checked against its cited evidence or
discarded. Pilot-eligible unresolved fields become work orders for approved pilot jobs.
Reconciliation then applies the source priority assigned to each field.}
  \label{fig:profile-pipeline}
  \vspace{-1.6em}
\end{figure*}

\subsection{Where That Knowledge Lives}
Consider what a login node reports. On TACC Stampede3~\cite{stampede3},
\texttt{sinfo -h -o "\%P \%G"} returns \texttt{(null)} for the \texttt{h100} and
\texttt{pvc} partitions, and \texttt{scontrol show node} shows a \texttt{CfgTRES}
of only CPU and memory. The scheduler tracks no GPU as a resource, 
yet the user guide lists four per node~\cite{stampede3-userguide}.
Enforced limits show a deeper gap. On Purdue Anvil~\cite{song2022anvil},
\texttt{sinfo} reports unlimited walltime, while the docs cap
\texttt{wholenode} jobs at 96 hours~\cite{anvil-userguide}. Some facts may 
not be documented clearly in either source. Whether a worker can
reach the outbound internet or access a particular storage directory may
therefore need to be established by running a pilot job that tests it.
Which source to trust differs from field to field. This is a crucial decision 
that requires reconciliation.

\subsection{The HPC Site Profile}
What deployment needs, then, is not simply a machine-readable description of a
site, but one that records how each value was established. We call this artifact
the \emph{HPC site profile}. Every field carries its evidence and provenance,
whether the value came from login-node measurement, documentation, or a pilot
job, and when it was observed. Absence is recorded rather than guessed. A field
that cannot be resolved is left unset and marked with the action that could
establish it. When sources disagree, the profile retains the conflicting
evidence and resolves the field in a separate reconciliation step.
Figure~\ref{fig:profile-excerpt} shows a profile excerpt beside its linked
evidence report. The GPU count, drawn from documentation, is kept over the
scheduler's conflicting null, and each value points to the measurement,
document, or pilot that supports it.

\section{Constructing the Profile}
\label{sec:construct}

\subsection{Overview and Trust Boundary}
A unified site profile is constructed for each site and reused across workflows
and users. Before any value is filled, a predefined rule assigns each field
its permitted evidence sources and their precedence:
login-node measurement for observable configuration
like scheduler type and storage paths (\texttt{measured\_config}), documentation
for policy limits and requirements (\texttt{documented\_policy}), and pilot
results for demonstrated compute-node behavior
(\texttt{demonstrated\_capability}). Construction then follows three steps,
probing the login node, extracting typed fields from documentation, and running
approved pilot jobs for eligible unresolved fields.
Figure~\ref{fig:profile-pipeline} shows this pipeline.

Only the documentation path uses a language model, and its authority is narrow.
It may select pages, propose queries, and propose typed fields with citations,
but cannot run commands, submit jobs, or write to the profile. Rule-based checks validate schema, site scope, and citation provenance, accepting or discarding each field, and unresolved fields remain unset. Everything outside the model's narrow authority is bounded or deterministic.

\subsection{Probing the Login Node}
The login-node probe runs a fixed set of reviewed commands. It records site
identity, the scheduler and its version, submit-command availability, reported
resources, and storage paths visible from the login node. These observations
populate \texttt{measured\_config} fields. The probe cannot establish
what a compute node sees or reaches, so those fields remain unset for pilot
jobs.

The measurements also seed documentation queries. Partition and filesystem
names are carried forward so documented values bind to known entities, while
measured values such as core counts, memory, and time limits are withheld from
extraction. This allows disagreements between measurement and documentation to
surface and be recorded rather than silently overwritten.

\subsection{Discovering Documentation}
\begin{figure}[t]
  \centering
  \includegraphics[width=0.47\textwidth]{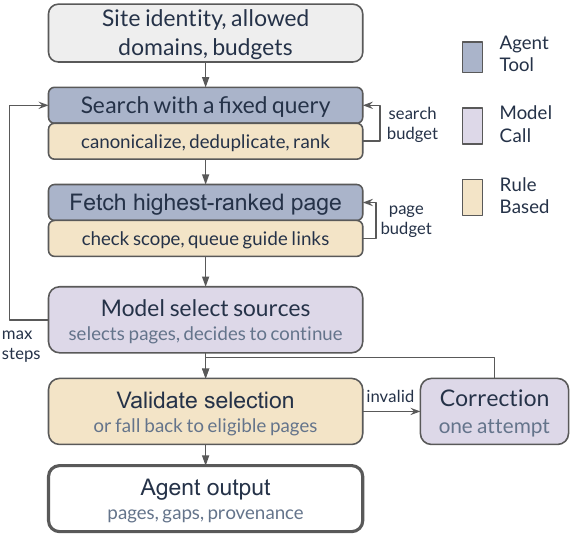}
  \caption{Bounded documentation discovery. Search results are canonicalized,
deduplicated, ranked, and restricted to allowed domains. The model selects
authoritative fetched pages and may request another bounded round.
Deterministic validation allows one correction attempt before final output.}
  \label{fig:discovery}
  \vspace{-1.6em}
\end{figure}

We use a single bounded agent to discover the site's documentation.
Figure~\ref{fig:discovery} shows the agentic loop. The agent begins with web
searches using queries derived from the site identity. Results are
canonicalized, deduplicated, ranked, and checked against an allowed-domain
list. It fetches the highest-ranked pages, verifies that they describe the
target system, and follows eligible links to other official guide pages. This
scope check prevents documentation for a sibling system at the same center
from entering the corpus.

After each search-and-fetch round, the model reviews only the fetched pages and
returns either \texttt{complete} or \texttt{search\_more}, along with selected
pages and unanswered topics. If more evidence is needed, it may propose up to
three additional queries, which pass through the same ranking, domain, and
scope checks. A \texttt{complete} decision ends discovery, while a fixed step
bound limits additional rounds.

The final selection is validated against a closed schema, with one correction
attempt and a fallback to eligible pages. Discovery terminates with the
selected pages, unresolved topics, and provenance for the searches and
documents that produced them.

\subsection{Corpus Construction and Retrieval}
The selected pages become a persistent corpus. Each document retains its URL,
title, fetch time, and content hash; text is split into bounded chunks along
the heading structure with tables kept intact; and individual sentences and
table rows carry their own span identifiers, which are the unit a citation may
point to. One frozen corpus is shared across all retrieval comparisons,
isolating retrieval from discovery.

Retrieval is lexical~\cite{robertson2009probabilistic} rather than dense, so
that identical queries against an identical corpus return identical chunks. We
compare three modes: full-corpus, which passes every target-site chunk in
bounded batches; BM25, which retrieves chunks per field using fixed queries
under a shared size budget; and LLM-expanded BM25, which adds at most two
model-generated variants per field and is the
only mode whose retrieved context may differ between runs.

\subsection{Typed Extraction and Evidence Validation}

The retrieved context is passed to the model as typed spans, individual
sentences or table rows with stable identifiers. The model receives those
spans and a closed output schema, and may return only typed field
candidates together with the identifiers of the spans that support each
value. 

Validation checks that every cited span exists, came from a chunk retrieved
for that field, belongs to the target site, and names a scheduler option,
partition, or storage resource admitted by the profile contract. Rejected
candidates may be repaired by one constrained correction call; what still
fails is discarded, and fields with no accepted candidate remain unresolved.
These checks establish schema, scope, and provenance consistency.
Whether the cited text supports the extracted value is evaluated against
hand-written ground truth in Section~\ref{sec:eval}, not inferred from citation
validity alone.

\begin{figure*}[t]
  \centering
  \includegraphics[width=0.9\textwidth]{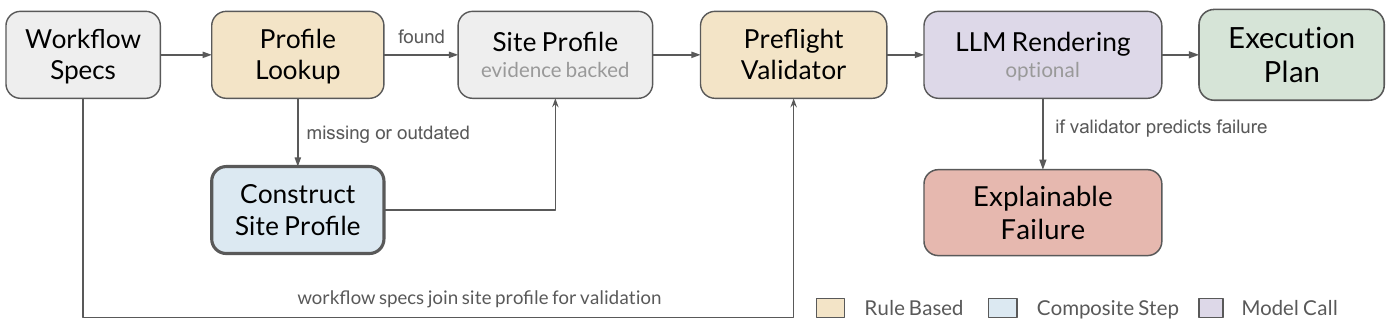}
  \caption{Preflight compares a normalized workflow specification with an
evidence-backed site profile. Compatible requirements yield a site-specific
execution plan, and a contradiction or unsatisfiable requirement yields an
explainable pre-submission failure.}
  \label{fig:preflight-pipeline}
  \vspace{-1.2em}
\end{figure*}

\subsection{Pilot Jobs and Reconciliation}
Among the unresolved fields, those with a registered pilot are mapped to fixed
jobs in a reviewed registry. Not every field can be resolved this way, and pilots
are used only for behavior that can be observed from a compute node. They test
whether a worker can read or write a storage location, reach the outbound
network, connect back to a manager, or communicate with another worker. A model
cannot create or modify a pilot, and live submission requires explicit
approval. 

Reconciliation then applies each field's predefined source precedence. Agreement
produces one value with all supporting evidence. If the highest-precedence source
is silent, a lower one may populate the field with its own provenance. When
sources disagree, the profile records the conflict and takes the value from the
highest-precedence source rather than the first one found. Fields no source
resolves are left unset.

\section{Workflow Preflight}
\label{sec:preflight}

Preflight consumes a normalized workflow specification that describes a
workflow's requirements independently of any particular workflow system. 
The specification is an extensible, machine-readable contract covering resource
shape, runtime limits, storage and network assumptions, and scheduler
constraints. A small workflow adapter translates each workflow system's native representation
into this common contract. We implement the first adapter for Floability
Backpacks, combining the backpack specification with command-line overrides to
recover the workflow's effective requirements~\cite{floability-backpack-escience25}.
Adapters for other workflow systems need only produce the same normalized
specification.

Given a target site, the system looks up a matching profile and checks its
schema and freshness. A missing or stale profile invokes the
construction process of Section~\ref{sec:construct}; a valid profile proceeds
directly to comparison, as shown in Figure~\ref{fig:preflight-pipeline}. Users configure when a field becomes stale and needs a recheck.

The validator compares each requirement against its profile field, distinguishing 
an incompatibility from missing evidence. A workflow exceeding a known limit is 
blocked; one requiring unresolved behavior returns unknown rather than being assumed 
safe or rejected. If all required fields are satisfied, the result is an execution 
plan with the selected resource, scheduler arguments, network settings, and a modified 
system command if needed. Otherwise each issue names the field, its required and observed 
values, and a remediation. Preflight never submits the workflow.

The structured result is authoritative and is produced without a language
model. An optional final model call may translate it into a human-readable
launch plan or failure explanation. The model receives only the completed
structured result and cannot change its status, selected resources, command,
or remediation. Without this call, the same decision and execution plan remain
available as structured JSON.

\section{Evaluation}
\label{sec:eval}

We evaluate the pipeline with three questions. \textbf{RQ1}: how accurately and conservatively does construction fill fields, and how often do its sources disagree? \textbf{RQ2}: how do model and retrieval choices affect profile quality, token use, latency, and variance? \textbf{RQ3}: can a profile preflight a real workflow into an execution plan or catch an incompatibility before submission?

\subsection{Experimental Setup}

\textbf{Sites.} We construct profiles for Purdue Anvil and TACC Stampede3,
which use Slurm, and the Notre Dame CRC, which uses HTCondor.

\textbf{Ground truth.} We hand-build one reference profile per site from
documentation, read-only login measurements, and pilot results. Each evaluated
field records its expected value or expected absence; uncertain fields are
excluded.

\textbf{Frozen inputs.} A preliminary run of 36 discovery builds with four models
across three sites returned the same authoritative
documentation in every run, independent of model. Since discovery added no variability, we
run it once per site and reuse the resulting corpus, along with the login
measurements and pilot results, across all runs. This isolates the effects of
model and retrieval choice for the extraction step. 

\textbf{Configurations.} We cross four language models, GPT-5 mini,
GPT-5.6 Terra, Gemini 3.6 Flash, and Gemini 3.1 Pro, with three retrieval
modes: BM25, LLM-expanded BM25, and full-corpus. Each site-model-retrieval
configuration is repeated three times, producing
$3 \times 4 \times 3 \times 3 = 108$ profile builds.

\textbf{Metrics.} A field is correct when its populated value, or its
absence, matches the reference profile. We report field accuracy over all
evaluated fields. Precision is measured over the fields a profile populates,
recall over the fields the reference fills, and abstention accuracy over the
fields the reference marks absent. Mechanical citation integrity is the
fraction of accepted citations whose cited chunk exists, carries target-site
scope, matches its source, and contains the quoted text. We also report the
rate at which a site's measured and documented sources disagree. Cost is
extraction tokens, latency, and run-to-run variation across each
configuration's three repetitions.

\subsection{RQ1: Profile Construction Quality}

RQ1 asks whether the three evidence paths and deterministic verification
produce accurate profiles without filling unsupported values, and how often
the sources disagree.

\begin{table}[t]
\caption{Profile-construction quality by site over 36 runs per site.}
\label{tab:rq1-site-results}
\centering
\footnotesize
\setlength{\tabcolsep}{2.5pt}
\begin{tabular}{lccccc}
\toprule
Site &
Precision &
Recall &
\shortstack{Abstention\\Accuracy} &
\shortstack{Field\\Accuracy} &
\shortstack{Citation\\Integrity} \\
\midrule
Anvil           & 0.979 & 0.874 & 0.969 & 0.895 & 1.000 \\
Stampede3       & 0.957 & 0.904 & 0.934 & 0.910 & 1.000 \\
Notre Dame CRC  & 0.992 & 0.771 & 0.986 & 0.783 & 1.000 \\
\bottomrule
\end{tabular}
\end{table}

Table~\ref{tab:rq1-site-results} shows high precision across all sites,
ranging from 0.957 to 0.992, while field accuracy ranges from 0.783 to 0.910.
Stampede3 achieves the highest field accuracy at 0.910, followed by Anvil at
0.895. Notre Dame CRC has the highest precision and abstention accuracy, but
its recall falls to 0.771. Its lower accuracy therefore results mainly from
fields remaining unresolved rather than from unsupported values entering the
profile. All 5,592 accepted citations pass the mechanical integrity checks.

\begin{figure}[t]
  \centering
  \includegraphics[width=0.98\columnwidth]
  {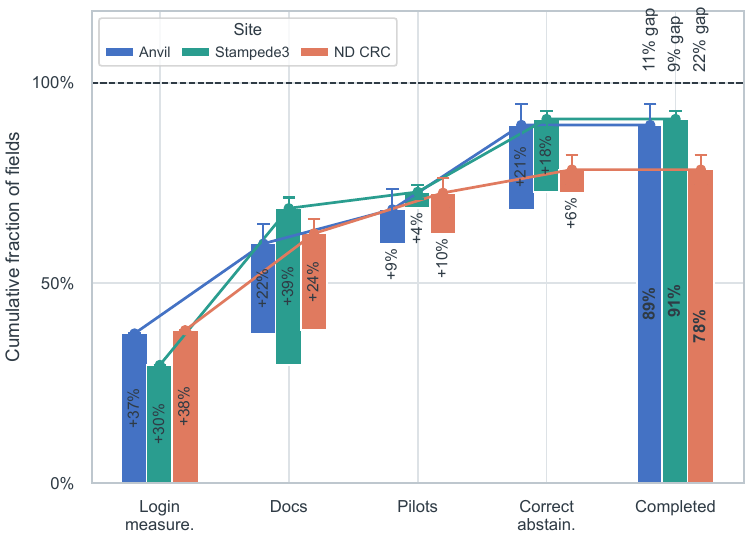}
  \caption{Profile completion by evidence path over 36 runs per site.
  Floating bars show the incremental fraction of fields resolved correctly by
  login measurement, documentation, pilot jobs, and correct
  abstention. The final bars show total field accuracy; the labeled gap is the
  fraction of incorrect or unresolved fields. Error bars show
  $\pm 1$ standard deviation of cumulative completion across runs.}

  \label{fig:profile-completion}
\end{figure}

Figure~\ref{fig:profile-completion} shows that the three evidence paths are
complementary. Login measurements resolve 30--38\% of the evaluated fields,
documentation adds 22--39\%, and pilot jobs add another 4--10\%. Together,
these sources resolve 68\% of Anvil, 73\% of Stampede3, and 72\% of Notre Dame
CRC before abstention is considered. Correct abstention raises field accuracy
to 89\%, 91\%, and 78\%, respectively. The remaining gaps of 9--22\% show that
deterministic verification keeps unsupported values out of the profile, but
does not eliminate missing site knowledge. Notre Dame CRC shows both sides of this: its 22\% gap is almost entirely abstention, and the per-PI storage of Section~\ref{sec:eval}-D is a documented
field that went unrecovered, which is what causes the false failure there.

Measurement and documentation also disagreed. Across the runs, we observed up
to eight conflicting fields at Anvil and six at Stampede3, with none at Notre
Dame CRC. Nearly all involved partition walltime caps for which the scheduler
reported no limit while the documentation stated one. Measurement was the more
permissive source in every case.

\subsection{RQ2: Model and Retrieval Effects}

RQ2 asks whether more capable models or more elaborate retrieval improve the
verified profile enough to justify their added cost.

\begin{table}[t]
\caption{Profile quality and extraction cost by model (27 runs per model).}
\label{tab:rq2-model-results}
\centering
\footnotesize
\setlength{\tabcolsep}{2pt}
\begin{tabular}{lcccccc}
\toprule
Model &
\shortstack{Field\\Accuracy} &
Precision &
Recall &
Tokens &
\shortstack{Latency\\(s)} &
\shortstack{Run\\SD} \\
\midrule
GPT-5 mini             & 0.841 & 0.959 & 0.826 & 89,513 & 362.0 & 0.040 \\
GPT-5.6 Terra          & 0.866 & 0.982 & 0.851 & 58,368 &  51.9 & 0.012 \\
Gemini 3.6 Flash       & 0.875 & 0.980 & 0.864 & 87,109 & 150.2 & 0.013 \\
Gemini 3.1 Pro         & 0.869 & 0.980 & 0.857 & 86,481 & 221.3 & 0.014 \\
\bottomrule
\end{tabular}
\end{table}

Table~\ref{tab:rq2-model-results} shows that deterministic verification narrows
the effect of model choice. Field accuracy ranges from 0.841 to 0.875, and the
three strongest models differ by less than one percentage point. Gemini 3.6
Flash achieves the highest accuracy at 0.875. GPT-5.6 Terra reaches 0.866 while
using the fewest tokens, completing fastest, and showing low run-to-run
variation. GPT-5 mini is the main outlier, with lower accuracy, higher latency,
and the largest variance.

\begin{table}[t]
\caption{Profile quality and extraction cost by retrieval strategy (36 runs each).}
\label{tab:rq2-retrieval-results}
\centering
\footnotesize
\setlength{\tabcolsep}{2pt}
\begin{tabular}{lcccccc}
\toprule
Retrieval &
\shortstack{Field\\Accuracy} &
Precision &
Recall &
Tokens &
\shortstack{Latency\\(s)} &
\shortstack{Run\\SD} \\
\midrule
BM25 &
0.868 & 0.976 & 0.855 & 57,590 & 129.9 & 0.021 \\
\shortstack[l]{LLM-exp. BM25} &
0.859 & 0.975 & 0.847 & 57,678 & 143.2 & 0.028 \\
Full corpus &
0.861 & 0.976 & 0.847 & 125,835 & 315.9 & 0.017 \\
\bottomrule
\end{tabular}
\end{table}

Retrieval choice affects cost far more than quality.
Table~\ref{tab:rq2-retrieval-results} shows that BM25 reaches the highest
retrieval-level accuracy, 0.868, using 57,590 tokens. LLM-expanded BM25 spends
nearly the same tokens for slightly lower accuracy, higher latency, and greater
variance. Full-corpus retrieval uses 2.18 times the tokens and 2.43 times the
latency of BM25 and still reaches lower accuracy.

\begin{figure*}[t]
  \centering
  \includegraphics[width=\textwidth]{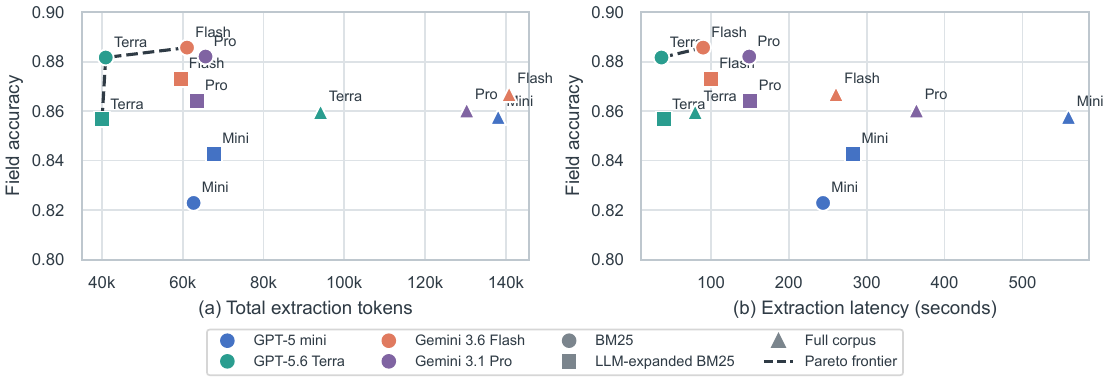}
  \caption{Quality--cost trade-offs by model and retrieval strategy, averaged
over nine runs per point. Panels compare field accuracy with provider-reported
tokens (a) and extraction latency (b). Color denotes model, shape denotes
retrieval, and the dashed line marks the observed Pareto frontier. The
accuracy axis is truncated to expose differences.}
  \label{fig:rq2-quality-cost}
  \vspace{-1em}
\end{figure*}

Figure~\ref{fig:rq2-quality-cost} plots both trade-offs. Terra with BM25 is the
lowest-cost high-accuracy configuration and Flash with BM25 is the most
accurate, and the two sit on the observed Pareto frontier while every
full-corpus point falls below and to the right of it. For this schema and
validation pipeline, BM25 is sufficient: neither full-corpus retrieval nor
model-generated query expansion improves the verified profile.

\subsection{RQ3: Preflight Case Study on a Real Workflow}
We evaluate preflight qualitatively through a case study, preflighting one
Floability Backpack against all three site profiles. The backpack requests one
to four workers, each with 112 CPU cores, two GPUs, and 128{,}000~MB of memory,
and its data sources require 108~GB to be staged before the run.

On Anvil the request was satisfied. Preflight matched it to the \texttt{gpu}
partition, which provides 128 CPU cores and four GPUs per node, selected the
site scratch filesystem for the 108~GB stage, and produced a Floability command
with the required CLI options and pilot-verified manager
ports. The result is an execution plan grounded in the profile.

On Stampede3 the request was correctly blocked. Its GPU partitions supply
enough GPUs and memory but only 96 CPU cores per node, short of the 112
requested, so preflight reported the unsatisfied field before submission rather
than after a queued job failed. 

On the Notre Dame CRC, the outcome was a false failure. Preflight produced
correct HTCondor syntax for the requested CPU and GPU nodes but blocked on
storage. CRC stages data through per-PI \texttt{/group} storage, which the
login probe missed, so the profile exposed only the home directory as a
possible staging location. The home directory lacked enough available space
for 108~GB of data, causing preflight to block a workflow the site could have
run. The error was in the profile, not the preflight logic, showing that
preflight is only as complete as its underlying profile.

\subsection{Qualitative Observations and Discussion}

Pilot jobs contribute the fewest fields, but establish compute-node behaviors
that neither measurement nor documentation can resolve. This supports the central profile design: useful site knowledge must combine multiple evidence paths, with deterministic verification deciding what may enter the profile.

Standard BM25 with fixed queries produced both the highest retrieval-level
accuracy and the lowest cost. Supplying the full corpus did not improve the
profile and sometimes produced worse extraction, possibly because irrelevant
or competing context obscured evidence for the closed, typed schema.
LLM-expanded queries provided no benefit either. These results may be specific
to a schema whose required fields are known in advance; broader retrieval or
query expansion could be more useful for open-ended profile construction. The
small differences among the stronger models further suggest that the main
benefit comes from constrained extraction and verification rather than model
choice.

The Notre Dame CRC preflight case also exposes an important limitation. Preflight
correctly applied the available profile but falsely blocked the workflow
because the profile omitted the site's per-PI storage. Thus, preflight is only
as reliable as the evidence represented beneath it. Conservative abstention
prevents unsupported plans, but it may also reject feasible deployments when
important site knowledge remains undiscovered. Deployment outcomes should
therefore be used to identify profile gaps and guide later measurements,
documentation searches, and pilot jobs.

\section{Related Work}
\label{sec:related}

\textbf{Packaging and portability.} A large body of work makes the software half
of a workflow portable. Package managers reconstruct software stacks from a
specification~\cite{gamblin2015spack, anaconda_conda_pack_2018}, container
runtimes ship a filesystem image~\cite{priedhorsky2017charliecloud,
kurtzer2017singularity}, and workflow standards describe the computation
independently of an execution engine~\cite{crusoe2022cwl}. Workflow engines and
job-submission APIs abstract the submit interface across schedulers, so one call
replaces a site-specific \texttt{qsub} or
\texttt{sbatch}~\cite{makeflow-sweet12, hategan2023psi, taskvine-works-2023,
babuji2019parsl}. Our own backpacks package a workflow for portable
deployment~\cite{floability-backpack-escience25}. All of these make a workflow
portable, yet each still leaves the user to supply the site-specific values a
submission needs, a gap the site profile fills.

\textbf{Machine-readable site descriptions.} The closest prior artifacts are
the Pegasus Site Catalog~\cite{deelman2015pegasus}, which records where data may
be staged and what the compute nodes look like, and grid information models such
as GLUE~\cite{andreetto2010standards}, which also model queue policies like walltime
and job-size limits. The site profile differs in three respects. It carries
requirement fields a workflow deployment needs beyond resource shape, such as
network reachability and node-local storage survival. It attaches provenance and
evidence to every field. And it reconciles a field across measurement,
documentation, and pilot sources rather than recording a single stated value.
The workflows community has repeatedly named multi-facility execution and
heterogeneous resources as open challenges~\cite{da2024workflows}.

\textbf{LLM agents over technical documentation.} Recent systems apply language
models to HPC user support and documentation. AskHPC and ChatHPC answer user 
questions by retrieving over fragmented
documentation~\cite{bondapalli2025askhpc,yin2025chathpc}. AskHPC identifies
the integration of live system information as future work. Such systems
produce an answer or a script for one request; none produces a reusable,
verified artifact, and none checks documentation against the running system,
which is where our configuration-versus-documentation disagreements arise. Our
agent follows the reason-and-act pattern~\cite{yao2022react} but is bounded: it
reads and proposes, while deterministic rules decide what enters the profile.

\textbf{Grounded extraction and citation.} Retrieval-augmented generation~\cite{lewis2020retrieval} and lexical
retrieval~\cite{robertson2009probabilistic} are the mechanisms we use for
context selection, while prior work studies attribution of generated claims to
supporting spans~\cite{gao2023enabling}. We adopt span-level attribution and
deterministically verify each citation against the frozen corpus. Accepted
documentation fields are then reconciled with measurement and pilot evidence
according to the source priority defined by the profile contract. This process
allows the profile to preserve and quantify disagreements between documented
and measured site information.

\section{Conclusions and Future Work}
\label{sec:conclusion}
The missing half of workflow portability is reliable knowledge of how an HPC
site must be used. We address this gap with an evidence-backed site profile
constructed from login-node measurements, authoritative documentation, and
approved pilot jobs. A bounded agent assists with documentation discovery and
typed extraction, while deterministic verification controls which fields enter
the profile. Across the evaluation, model choice had little effect after
verification, and standard BM25 achieved comparable quality at lower cost than
more expensive retrieval strategies. The resulting profiles supported 
site-specific preflight, correctly rejected
an incompatible resource request, and exposed how missing profile evidence can
also cause a false failure.

This paper establishes the site profile concept and implements its core mechanisms, 
while leaving broader coverage for future work. Preflight was evaluated only with a
Floability Backpack, and the current profile cannot guarantee that it captures
all information required by other workflow systems. Future work
will extend the schema, measurements, and pilot coverage for additional workflow
systems and scheduler environments. We will also explore how to incorporate
site-specific best practice, alongside the hard
constraints, and move from preflight to direct deployment. Finally, we plan to
feed observed deployment successes, failures, and runtime behavior back into
subsequent profile construction and planning.

\section*{Availability}
The open-source implementation, schemas, site examples,
and detailed results from the 108-run evaluation are available at
\url{https://github.com/floability/hpc-site-preflight}.

\section*{Acknowledgment}
This project is supported by the U.S.\ National Science Foundation under grant OAC-2411436.

\bibliographystyle{IEEEtran}
\bibliography{references}
\end{document}